\documentstyle[12pt]{article}
\makeatletter
\ifcase \@ptsize
\or 
\or 
\fi
\advance\textheight by \topskip

\ifcase \@ptsize
\or 
\or 
\fi

\def\@cite#1#2{\nolinebreak$^{[\scriptstyle #1\if@tempswa , #2\fi]}$}
\def\@citex[#1]#2{\if@filesw\immediate\write\@auxout{\string\citation{#2}}\fi
  \def\@citea{}\@cite{\@for\@citeb:=#2\do
    {\@citea\def\@citea{,\penalty\@m}\@ifundefined
       {b@\@citeb}{{\bf ?}\@warning
       {Citation `\@citeb' on page \thepage \space undefined}}%
{\csname b@\@citeb\endcsname}}}{#1}}

\@definecounter{footnote}

\def\section{\@startsection {section}{1}{\z@}{-3.5ex plus -1ex minus
 -.2ex}{2.3ex plus .2ex}{\large\bf\centering}}
\def\subsection{\@startsection{subsection}{2}{\z@}{-3.25ex plus -1ex minus
 -.2ex}{1.5ex plus .2ex}{\sc}}

\begin{document}
\baselineskip 18pt
\parindent 15pt
\parskip 0.2in
\newcommand{\hsp}{\hspace{0.08in}}
\newcommand{\A}{{\cal A}}
\centerline{\large On the bootstrap structure of Yangian-invariant}
\centerline{\large factorized S-matrices}

\vspace{0.2in}
\centerline{Niall J. MacKay\footnote{Supported by a UK SERC studentship.}}

\vspace{0.1in}
\centerline{\em Department of Mathematical Sciences, Durham, DH1 3LE, England}

\vspace{0.2in}
The message of this talk is that there is some very rich,
largely unremarked structure in the representation theory
of the Yangian (a quasitriangular Hopf algebra discovered
by Drinfeld\cite{D} in 1985).

We begin by taking the point of view\cite{L,Bern} of a physicist
investigating the charge algebra of integrable field theories in
1+1 dimensions which have a Lie algebra valued, curvature-free,
conserved current (of which notable examples
are the principal chiral model and the Gross-Neveu model and its
generalizations). Such theories thus contain a charge $Q_0^a$,
valued in a Lie algebra $\A$, which acts additively on tensor
products of asymptotically independent particle states: defining
its action on a two-particle state to be $\Delta(Q_0^a)$, we have
\begin{equation}\label{cr0}
\left[ Q_0^a , Q_0^b \right] = i \hbar f^{abc} Q_0^c
\hspace{0.2in}{\rm and}\hspace{0.2in}
\Delta(Q_0^a) = Q_0^a \otimes 1 + 1 \otimes Q_0^a \hsp.
\end{equation}
Such theories also have higher {\em non-local} conserved charges.
The first of these, $Q_1^a$, satisfies
\begin{equation}\label{cr1}
\left[ Q_0^a , Q_1^b \right] = i \hbar f^{abc} Q_1^c
\hspace{0.2in}{\rm and}\hspace{0.2in}
\Delta(Q_1^a) = Q_1^a \otimes 1 + 1 \otimes Q_1^a +
{1\over 2}f^{abc} Q_0^c \otimes Q_0^b \hsp.
\end{equation}
The missing commutator of two $Q_1$'s is fixed by requiring
that asymptotic states carry a representation of the charge
algebra ({\em i.e.\ }that $\Delta$ be a homomorphism), and is ($\A\neq sl(2)$)
\begin{equation}\label{YSerre}
 f^{d[ab} [Q_1^{c]},Q_1^d] \hsp = \hsp {i\hbar\over{12}}
\,f^{ap i} f^{bq j}f^{cr k}f^{ijk} \, Q_0^{(p} Q_0^q Q_0^{r)} \hsp.
\end{equation}
This complicated relation may safely be ignored for the moment,
 although we shall return to it later.
The Lorentz boost does not act trivially on these charges. If
we boost the rapidity
(defined by ${\bf p}=(m\,{\rm{cosh}}\theta,m{\rm\,{sinh}}\theta)$)
of a particle by $\theta$, it is found that
\begin{equation}\label{Lorentz}
L_\theta : \hspace{0.1in} Q^{(0)} \mapsto Q^{(0)} \hsp,
\hspace{0.2in}Q^{(1)} \mapsto Q^{(1)} - {{\hbar C_{Adj}}\over{4\pi}}
\, \theta \, Q^{(0)} \hspace{0.25in}{\rm (where}\hspace{0.1in}
C_{Adj}\delta_{ad} = f^{abc}f^{cbd}\hsp{\rm ).}
\end{equation}
This has important implications for the $S$-matrix, which is
heavily constrained by conservation of $Q_0$ and $Q_1$. The
$S$-matrix for the interaction of two particles of rapidities
$\theta_1$ and $\theta_2$ must satisfy
\begin{eqnarray}\nonumber
 & \left[ \,S(\theta_1-\theta_2) \, , \, Q_0^a \otimes 1 +
1 \otimes Q_0^a \,\right] = 0 & \\[0.1in] \label{sm}
 & S(\theta_1-\theta_2) \left( L_{\theta_1} \otimes L_{\theta_2}
\Delta(Q_1^a) \right) = \left( L_{\theta_2} \otimes L_{\theta_1}
\Delta(Q_1^a) \right) S(\theta_1-\theta_2) \hsp. &
\end{eqnarray}
These equations, together with unitarity and crossing-symmetry,
determine the $S$-matrix up to an overall scalar factor.
The classic result implied by these equations is that the
$S$-matrix is {\em factorized}: that the $S$-matrix for the
interaction of a large number of particles may be written as a
product of two-particle $S$-matrices. That this works consistently
follows from the Yang-Baxter equation (YBE),
\begin{equation}\label{ybe}
S_{V_2V_3}(\theta')S_{V_1V_3}(\theta+\theta')S_{V_1V_2}(\theta)
= S_{V_1V_2}(\theta)S_{V_1V_3}(\theta+\theta')S_{V_2V_3}(\theta')
\end{equation}
(where $\theta$ and $\theta'$ denote the appropriate rapidity
differences), which follows from (\ref{sm}): the $S$-matrix does
not depend on the order in which the particles interact, and its
only action is to exchange quantum numbers within the multiplets
$V_i$ of particles of equal mass $m_i$.
Factorization was actually deduced some years ago from other
considerations, and such $S$-matrices have been studied for
some time\cite{ZZ,OW}. In fact, as Bernard\cite{Bern} pointed out,
the algebra (\ref{cr0},\ref{cr1},\ref{YSerre}) is precisely
Drinfeld's Yangian $Y(\A)$, which is a quasitriangular Hopf
algebra
with an $R$-matrix satisfying (\ref{sm},\ref{ybe}) (where
$S={\bf P}R$ with ${\bf P}$ the transposition map), and has
a one-parameter automorphism (\ref{Lorentz}).
Now suppose that we wish to construct $S$-matrices without
using the underlying Yangian structure. In order to do so
we may use the {\em bootstrap principle}: that, at appropriate
 poles, intermediate states of the $S$-matrix should be regarded
as physical particles. For example, where ${\bf 1}$ is the particle
in the vector
representation of $SO(N)$ we have\cite{ZZ}
$$
S_{\bf 11}(\theta) = s(\theta) \left[ (\theta_0 -\theta)(i\pi-\theta)
P_S +
(\theta_0 + \theta)(i\pi-\theta)P_A + (\theta_0 + \theta)(i\pi-\theta)
P_0 \right]
$$
where $P_S,P_A$ and $P_0$ project onto the traceless symmetric,
antisymmetric and trace components of ${\bf 1}\otimes {\bf 1}$,
$\theta_0={2i\pi\over{N-2}}$, and $s(\theta)$ is a scalar function.
The only value of $\theta$ in the physical strip $0 \leq {\rm Im}
\theta < \pi$ at which any of these components vanish is
$\theta=\theta_0$, and we therefore choose the ambiguous factor
$s(\theta)$ to have a simple pole at $\theta_0$, and interpret
$P_A+P_0$ as a particle state ${\bf 2}$, so that we have a
`fusing' ${\bf 11}\rightarrow{\bf 2}$. Conservation of momentum
tells us the relative masses of these particles, and we can
construct $S_{\bf 12}$ and $S_{\bf 22}$ according to the
principle\cite{Kar}

\hspace{10mm}
\setlength{\unitlength}{1mm}
\begin{picture}(155,40)(-65,-25)
\put(-30,-5){\line(-4,-5){16}}
\put(-30,-5){\line(2,-5){8}}
\put(-30,-5){\line(1,6){3}}
\put(30,-5){\line(-4,-5){16}}
\put(30,-5){\line(2,-5){8}}
\put(30,-5){\line(1,6){3}}
\put(-50,-20){\line(6,1){34}}
\put(20,0){\line(6,1){24}}
\put(-5,-3){\makebox(0,0){=}}
\put(-45,-28){\makebox(0,0){\bf 1}}
\put(-22,-28){\makebox(0,0){\bf 1}}
\put(15,-28){\makebox(0,0){\bf 1}}
\put(38,-28){\makebox(0,0){\bf 1}}
\put(35,13){\makebox(0,0){\bf 2}}
\put(-25,13){\makebox(0,0){\bf 2}}
\put(-58,-20){\makebox(0,0){{\bf 1} or {\bf 2}}}
\put(12,0){\makebox(0,0){{\bf 1} or {\bf 2}}}
\put(-30,-10){\makebox(0,0){$\theta_0$}}
\put(30,-10){\makebox(0,0){$\theta_0$}}

\end{picture}

We can now continue the bootstrap, examining $S_{\bf 12}$ and
$S_{\bf 22}$ for poles, interpreting these as third and higher
particles, and so on. In principle, we would hope that eventually
all appropriate poles would have been interpreted as physical
states, so that the bootstrap would have closed on a finite set
of particles. The mass ratios of the particles would be known,
as would the representation (possibly reducible) of $\A$ carried
by the particle, and the set of particle `fusings' ${\bf ab}\rightarrow
{\bf c}$.
Unfortunately, calculating the $S$-matrices involves a great deal
of computation, and calculations beyond $S_{\bf 22}$ are unfeasible
in practice.
However, we can already extract one important point. In all cases
hitherto computed, it has been possible to assign a particle {\bf i}
to a reducible representation of $\A$ containing the $i$th fundamental
representation $V_i$ of $\A$. In some cases ({\em e.g.\ }for $\A=a_n$
and for vector and spinor representations of $SO(N)$) there are no
other components. A
general method exists for calculating the $S$-matrices in irreducible
representations\cite{KSR}, and the bulk of the $S$-matrices so far
constructed have been of this type\cite{OW,KSR}, and led to the
conjecture that the fusings are in precise agreement with the
Clebsch-Gordan (CG) decomposition, {\em i.e.\ }that
${\bf ab}\rightarrow{\bf c}$ if and only if $V_a \otimes V_b
\supset V_c$. However, when the particle multiplet is reducible,
this fails: for example, when we construct $S_{\bf 22}$ for the
 $SO(N)$ theories\cite{facsm}, we find that ${\bf 22} \not\rightarrow
{\bf 2}$ in general even though $V_2 \otimes V_2 \supset V_2$.

Now let us compare these $S$-matrices with another type of
factorized $S$-matrix.
In {\em purely elastic scattering theories} or PESTs (of which
the prime examples are affine Toda theories\cite{BCDS2} and
integrable deformations of CFTs\cite{Z}), there is no
mass-degeneracy: the particles do not form multiplets, and
(\ref{ybe}) is trivial. The $S$-matrix is a scalar factor
determined by unitarity, analyticity, crossing-symmetry and the
bootstrap, which is now much easier to implement. The result is
a closed bootstrap on a full spectrum of massive particles and
their fusings. For  simply-laced $\A$ these match precisely the
known results for the YBE-dependent $S$-matrices (including the
CG `holes')\footnote{For non-simply-laced $\A$ there does seem
to be a relationship, but it is much less clear.}; moreover,
the PEST structure has a beautiful description in terms of root
systems of Lie algebras\cite{ped}. It therefore seems likely that
similar structure exists in the YBE-dependent $S$-matrices - it
is simply that the bootstrap, whose implementation is specific
to the $\A$ under consideration, is unable to perceive it.

Let us therefore shift viewpoint and consider the particle
multiplets {\bf i} as fundamental representations of $Y(\A)$.
We can use (\ref{Lorentz}) to boost ${\bf i}\mapsto {\bf i}(\theta)$;
the fusion procedure for $S_{\bf ab}(\theta)$ is then that of
decomposing ${\bf a}(\theta) \otimes {\bf b}$ into its
$Y(\A)$-irreducible components. An alternative approach to
constructing $S$-matrices is to construct the action of $Q_0$
and $Q_1$ explicitly and solve (\ref{sm}). This has been done,
 as far as I am aware, for two cases only. The first is when
${\bf i}=V_i$ is irreducible and $\rho_{\bf i}(Q_1)=0$: this
only works when the action of the right-hand side of (\ref{YSerre})
is zero.
The second is the extension of the adjoint representation of $\A$
by a scalar representation to the $Y(\A)$-irreducible representation
$\A \oplus {\bf C}$: the action of $Q_0$ is the adjoint action on
$\A$ and zero on ${\bf C}$, and the action of $Q_1$ is then
constructed to satisfy (\ref{cr1},\ref{YSerre}). Both of these
types of representation were discovered by Drinfeld\cite{D},
and the corresponding $R$-matrices have been calculated\cite{KSR,CP3}.
Unfortunately this is where the story currently ends.
The construction of other representations of $Y(\A)$ seems
to me to be an interesting problem for mathematicians: for
example, can we construct $\rho_{\bf i}$ by taking $Q_0$ to
have the natural action, and then fixing the action of $Q_1$
so that (\ref{cr1},\ref{YSerre}) are satisfied?
Beyond this, solving (\ref{sm}) for $S$ is certain to be a
tricky process, and may yield little insight into the information
obtained; at all levels, it seems, generality is lacking.

In conclusion, there is every prospect of interesting structure
in the representation theory of $Y(\A)$ alongside a dearth of
general techniques for its discovery. For mathematical physicists,
I should perhaps point out in addition the wide range of 1+1 dimensional
integrable theories to which $Y(\A)$ is relevant - apart from those
mentioned at the beginning, the close relationship with PESTs has
even led Belavin\cite{Bel} to suggest that the underlying symmetry
of affine Toda theories may be $Y(\A)/\A$. I can only echo this,
and hope to have convinced
people that the subject deserves more attention than it has
hitherto received.

\bibliographystyle{npb}

\begin{thebibliography}{10}

\bibitem{D}
V.~G. Drinfeld,
\newblock Sov.Math.Dokl. 32 (1985) 254.

\bibitem{L}
M.~L\"uscher,
\newblock Nucl.Phys. B135 (1978) 1.

\bibitem{Bern}
D.~Bernard,
\newblock Comm.Math.Phys. 137 (1991) 191.

\bibitem{ZZ}
A.~B. Zamolodchikov and A.~B. Zamolodchikov,
\newblock Ann.Phys. 120 (1979) 253.

\bibitem{OW}
E.~Ogievetsky and P.~Wiegmann,
\newblock Phys.Lett. B168 (1986) 360.

\bibitem{Kar}
M.~Karowski,
\newblock Nucl.Phys. B153 (1979) 244.

\bibitem{KSR}
P.~P. Kulish, E.~K. Sklyanin and N.~Y. Reshetikhin,
\newblock Lett.Math.Phys. 5 (1981) 393.

\bibitem{facsm}
N.~J. Mac{K}ay,
\newblock Nucl.Phys. B356 (1991) 729.

\bibitem{BCDS2}
H.~W. Braden, E.~Corrigan, P.~E. Dorey and R.~Sasaki,
\newblock Nucl.Phys. B338 (1990) 689.

\bibitem{Z}
A.~B. Zamolodchikov,
\newblock Int.J.Mod.Phys A4 (1989) 4235.

\bibitem{ped}
P.~E. Dorey,
\newblock Nucl.Phys. B358 (1991) 654.

\bibitem{CP3}
V.~Chari and A.~Pressley,
\newblock J.{f\"ur} die reine und ang. Math. 417 (1991) 87.

\bibitem{Bel}
A.~A. Belavin,
\newblock Phys.Lett. 283B (1992) 67.

\end{thebibliography}

\baselineskip 12pt
\small{

}
\end{document}